\documentclass[apj]{emulateapj}
\usepackage{amsmath}
\usepackage{apjfonts,lscape}

\newcommand{\czhel}{\mbox{$cz_{\rm hel}$}}	% Heliocentric radial velocity
\newcommand{\czsyshel}{\mbox{$cz_{\rm sys,hel}$}}
\newcommand{\czlg}{\mbox{$cz_{_{\rm LG}}$}}	% Velocity rel. to Local Group
\newcommand{\dmfifteen}{\mbox{$\Delta m_{15}$}} % Delta m_15
\newcommand{\dflow}{\mbox{$D_{\rm flow}$}}	% Distance from flow model
\newcommand{\dsnia}{\mbox{$D_{\rm SN\,Ia}$}}	% Distance by SN Ia method
\newcommand{\DIa}{\ensuremath{22.3 \pm 2.8}}	% Value of SN Ia distance
\newcommand{\dmodsnia}{\mbox{$\mu_{\rm SN\,Ia}$}}
\newcommand{\dtrgb}{\mbox{$D_{\rm TRGB}$}}	% Distance by TRGB method
\newcommand{\dmodtrgb}{\mbox{$\mu_{\rm TRGB}$}}	% Dist. modulus by TRGB method
\newcommand{\hi}{\ion{H}{1}}			% Symbol for neutral hydrogen
\newcommand{\hii}{\ion{H}{2}}			% Symbol for H~II region
\newcommand{\hst}{{\em HST}}			% Abbrev. for Hubble Space Tel.
\newcommand{\kms}{km~s$^{-1}$}			% kilometers per second
\newcommand{\kmsmpc}{km~s$^{-1}$~Mpc$^{-1}$}	% km/sec/Mpc
\newcommand{\n}{NGC~}				% NGC = New General Catalogue
\newcommand{\rvhost}{\mbox{$R_V^{\rm host}$}}	% Ratio R_V for host galaxy
\newcommand{\rvmw}{\mbox{$R_V^{\rm MW}$}}	% Ratio R_V for Milky Way
\newcommand{\sigvelth}{\mbox{$\sigma_{\rm th}$}} % Thermal velocity dispersion
\newcommand{\sneia}{SNe~Ia}			% Supernovae of Type Ia
\newcommand{\snia}{SN~Ia}			% Supernova of Type Ia
\newcommand{\trgbmag}{\mbox{$T_{\rm RGB}$}}	% T magnitude for RGB
\slugcomment{To appear in {\em The Astronomical Journal}, Vol.\ 136 (2008)}

\shorttitle{New Distance to The Antennae}
\shortauthors{Schweizer et al.}

\begin{document}

\title{A New Distance to The Antennae Galaxies (NGC 4038/39) Based on the Type I\lowercase{a}\\
	Supernova 2007\lowercase{sr}\altaffilmark{1}}

\author{
Fran\c cois Schweizer\altaffilmark{2},
Christopher R. Burns\altaffilmark{2},
Barry F. Madore\altaffilmark{2},
Violet A. Mager\altaffilmark{2},
M. M. Phillips\altaffilmark{3},\\
Wendy L. Freedman\altaffilmark{2},
Luis Boldt\altaffilmark{3},
Carlos Contreras\altaffilmark{3},
Gaston Folatelli\altaffilmark{4},
Sergio Gonz\'alez\altaffilmark{3},\\
Mario Hamuy\altaffilmark{4},
Wojtek Krzeminski\altaffilmark{3},
Nidia I. Morrell\altaffilmark{3},
S. E. Persson\altaffilmark{2},\\
Miguel R. Roth\altaffilmark{3},
and
Maximilian D.\ Stritzinger\altaffilmark{3}
}

\altaffiltext{1}{Based in part on observations with the 6.5 m Magellan
Telescopes located at Las Campanas Observatory, Chile.}
\altaffiltext{2}{Carnegie Observatories, 813 Santa Barbara Street, Pasadena,
   CA 91101; schweizer@ociw.edu; cburns@ociw.edu; barry@ociw.edu;
   vmager@ociw.edu; wendy@ociw.edu; persson@ociw.edu.}
\altaffiltext{3}{Las Campanas Observatory, Carnegie Observatories,
   Casilla 601, La Serena, Chile; mmp@lco.cl; lboldt@lco.cl;
   ccontreras@lco.cl; sgonzalez@lco.cl; wkrzeminski@lco.cl; nmorrell@lco.cl; 
   miguel@lco.cl; mstritzinger@lco.cl.}
\altaffiltext{4}{Universidad de Chile, Departamento de Astronom\'ia,
   Santiago, Chile; gfolatelli@lco.cl;  mhamuy@das.uchile.cl.}

% Notice that each of these authors has alternate affiliations, which
% are identified by the \altaffilmark after each name.  The actual alternate
% affiliation information is typeset in footnotes at the bottom of the
% first page, and the text itself is specified in \altaffiltext commands.
% There is a separate \altaffiltext for each alternate affiliation
% indicated above.

% The abstract environment prints out the receipt and acceptance dates
% if they are relevant for the journal style.  For the aasms style, they
% will print out as horizontal rules for the editorial staff to type
% on, so long as the author does not include \received and \accepted
% commands.  This should not be done, since \received and \accepted dates
% are not known to the author.

\begin{abstract}

Traditionally, the distance to \n4038/39 has been derived from the systemic
recession velocity, yielding about 20 Mpc for $H_0 = 72$ \kmsmpc.
Recently, this widely adopted distance has been challenged based on
photometry of the presumed tip of the red giant branch (TRGB), which
seems to yield a shorter distance of $13.3\pm 1.0$ Mpc and, with it,
nearly 1 mag lower luminosities and smaller radii for objects in this
prototypical merger.
Here we present a new distance estimate based on observations of the
Type Ia supernova (SN) 2007sr in the southern tail, made at Las Campanas
Observatory as part of the Carnegie Supernova Project.
The resulting distance of $\dsnia = \DIa$ Mpc
[$(m-M)_0 = 31.74\pm 0.27$ mag] is in good agreement with a refined
distance estimate based on the recession velocity and the large-scale flow
model developed by Tonry and collaborators, $\dflow = 22.5\pm 2.8$ Mpc.
We point out three serious problems that a short distance of 13.3 Mpc
would entail, and trace the claimed short distance to a likely
misidentification of the TRGB.
Reanalyzing {\em Hubble Space Telescope} (\hst\,) data in the Archive
with an improved method, we find a TRGB fainter by 0.9 mag and derive
from it a preliminary new TRGB distance of $\dtrgb = 20.0\pm 1.6$ Mpc.
Finally, assessing our three distance estimates we recommend using a
conservative, rounded value of $D = 22\pm 3$ Mpc as the best currently
available distance to The Antennae.

\end{abstract}

% The different journals have different requirements for keywords.  The
% keywords.apj file, found on aas.org in the pubs/aastex-misc directory, 
% contains a list of keywords used with the ApJ and Letters.  These are 
% usually assigned by the editor, but authors may include them in their 
% manuscripts if they wish. 


%\newpage
\section{INTRODUCTION}
\label{sec1}

The Antennae (\n4038/39) are the nearest example of a major merger
involving two gas-rich disk galaxies of comparable mass.
They allow us to study processes of dissipational galaxy assembly from
close up, thus providing a valuable glimpse of what must have been
more frequent events in the early universe.
The Antennae have been observed extensively at all wavelengths (e.g.,
X-rays: \citealt{fabb04}; UV: \citealt{hibb05}; optical: \citealt{whit99};
IR: \citealt{gilb07}; and 21-cm line: \citealt{hibb01}) and have also been
repeatedly modeled via N-body and hydrodynamical simulations
\citep[e.g.,][]{tt72,barn88,miho93,hibb03}.
An accurate distance to this prototypical merger is of obvious importance
in assessing its structure and dynamics, as well as for determining the
physical properties of its myriad of stars, star clusters, peculiar
objects such as ULXs (ultraluminous X-ray sources), and gas clouds.

Traditionally, the distance to \n4038/39 has most often been derived from
the systemic recession velocity and an adopted Hubble constant $H_0$,
with or without corrections for deviations of the Hubble flow from
linearity due to various attractors.
A frequently used modern value for the distance is 19.2 Mpc \citep{whit99}.
This value is based on a systemic recession velocity relative to the
Local Group of $\czlg = 1439$ \kms \citep{ws95,rc3}, a linear Hubble flow,
and $H_0 = 75$ \kmsmpc.
Updated to the \hst\ Key Project's derived $H_0 = 72\pm 8$ \kmsmpc\
\citep{freed01}, this often-used distance to The Antennae becomes
$20.0_{-2.0}^{+2.5}$ Mpc. 

A significantly shorter distance of $13.3\pm 1.0$ Mpc has recently been
determined by \cite{savi08} via photometry of the
% tip of the red giant branch (hereafter TRGB)
TRGB in a region near the tip of the southern tidal tail.
This photometry---performed on new, deep images obtained with \hst\,'s
Advanced Camera for Surveys (ACS)---seems to support the short distance
derived earlier from \hst/WFPC2 images of the same region by the same
method \citep{savi04}.
If the short distance is correct, The Antennae ``diminish'' in physical
size, mass, and luminosity, as do their stars, clusters, and gas clouds.
However, their recession velocity then deviates from the best
large-scale-flow model \citep[hereafter TBAD00]{tonr00} by about 500 \kms\
or 2.7\,$\sigma$ \citep{savi08}.

Clearly, new measurements of the distance to The Antennae by independent
methods are highly desirable.
In the present paper we derive a new distance from observations of
the Type Ia SN 2007sr, which appeared in the midsection of the southern
tail in 2007 December.
Section~\ref{sec2} presents our observations, data analysis, and new
distance.
Section~\ref{sec3} discusses problems with the short distance and
points out a likely error in the identification of the true TRGB
by \citet{savi08}.
Finally, \S~\ref{sec4} presents our conclusions and recommendation.

\begin{figure*}
  \centering
  \includegraphics[scale=0.545]{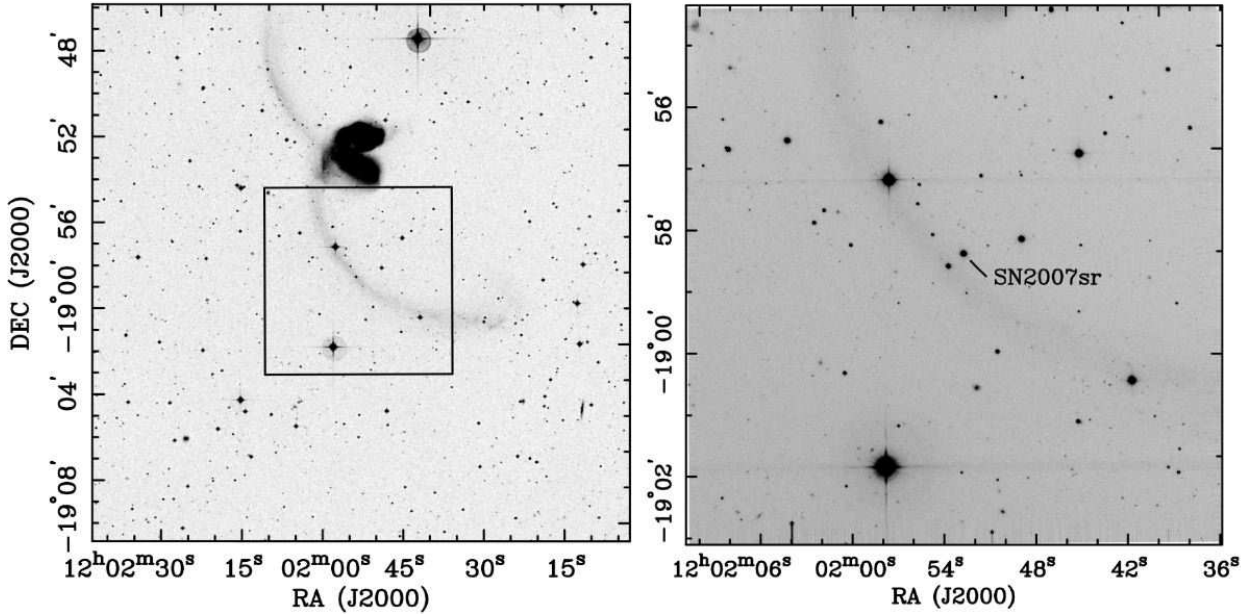}
  \caption{
Supernova 2007sr in the southern tidal tail of \n4038.
{\em Left:} Image of The Antennae reproduced from the Digital Sky Survey; the
box marks the field of view corresponding to the right panel.
{\em Right:} Image of SN 2007sr in the $V$ passband, produced by stacking
29 exposures of 50~s duration each obtained over many nights with the
CCD camera of the Swope 1.0~m telescope at Las Campanas Observatory.
  \label{fig01}}
\end{figure*}

%\bigskip					%% Remove before submission
\section{A NEW DISTANCE FROM SN 2007\lowercase{sr}}
\label{sec2}

Supernova 2007sr was discovered by the Catalina Sky Survey on
2007 December 18.53 UT \citep{drak07}, about 4.2 days after $B$ maximum
\citep{umbr07,pojm08}.
Since it was of Type Ia \citep{nait07,umbr07}, it was followed by the Carnegie
Supernova Project (CSP) as part of their low-z campaign \citep{Hamuy2006},
with observations beginning on 2007 December 20.33 UT.
Figure~\ref{fig01} shows the SN on a stacked $V$ image obtained with the
Swope 1~m telescope at Las Campanas Observatory.

The CSP measured light curves for the SN using the filter set $u'g'r'i'BVJH$
from 6 days until approximately 130 days after $B$ maximum.
Appendix~\ref{appA} gives some details about the photometric system and
calibration, while its Table~\ref{tab:photometry} lists all measured
magnitudes and their uncertainties.

It is now well known that supernovae of Type Ia (\snia) can serve as
excellent standard candles and have been used to both discover the presence
\citep{Riess1998,Perlmutter1999} and constrain the nature of 
dark energy \citep{Knop2003,Astier2006,Clocchiatti2006,WoodVasey2007}.
Once corrected for extinction and the well-known correlation between
intrinsic luminosity and decline rate of the light curve 
\citep{Phillips1993,Hamuy1996a,Riess1996},
\sneia\ have an intrinsic dispersion of approximately $\pm 0.15$ mag.
We can therefore use SN 2007sr to constrain the distance to \n4038/39 to
this level.

Because SN 2007sr was caught after maximum, it is necessary to use
\snia\ light-curve templates to estimate the peak magnitude in each passband
as well as the decline-rate parameter \dmfifteen, defined as
the change in $B$ magnitude between maximum and 15 days later in the
rest frame of the SN. This parameter will allow us to correct for
the intrinsic luminosity of the SN and also to predict its
intrinsic colors, from which we can derive the host galaxy extinction.
More formally, we model the light curve in each passband $X$ with the
following formula:
\begin{equation}
\begin{split}
   m_{X}=~&T_{X}\left(t-t_{\rm max},\dmfifteen\right)+M_{X}^{0}+
                  b_{X}\left(\dmfifteen-1.1\right)+   \\
          &R_{X}^{\rm MW}E(B-V)_{\rm MW}+R_{X}^{\rm host}E(B-V)_{\rm host}+
                  K_{X}+\mu_0\,,    \\
   \label{eq:lc_model}
\end{split}
\end{equation}
where $T_{X}$ is the light-curve template for passband $X$, $t_{\rm max}$
is the time of maximum for the $B$ light curve, $M_{X}^{0}$ is the absolute
magnitude in passband $X$ of a \snia\ with $\dmfifteen=1.1$ mag, $b_{X}$
is the slope of the luminosity--\dmfifteen\ relationship, $K_X$ is the
$K$-correction in passband $X$, and $\mu_0$ is the true distance modulus.
The extinctions from the Milky Way and host galaxy are
$R_{X}^{\rm MW}E(B-V)_{\rm MW}$
and $R_{X}^{\rm host}E(B-V)_{\rm host}$, respectively.
The color excess due to the Milky Way foreground extinction is taken
from \citet{schl98}, which for the position of SN 2007sr corresponds to 
$E(B-V)_{\rm MW}=0.046\pm0.005$ mag.

The values of the parameters $M_{X}^{0}$ and $b_{X}$ are determined from
the set of \sneia\ observed by the CSP in the first-year campaign by
assuming a fixed cosmological model ($H_0=72$ \kmsmpc, $\Omega_{\rm m}=0.28$,
and $\Omega_{\Lambda}=0.72$; \citealt{sper07}), computing distance moduli
from the observed redshifts of the host galaxies, correcting for Milky-Way
and host-galaxy extinction, and fitting the resulting absolute peak
magnitudes in each passband to the formula
\begin{equation}
   M_{X}=M_{X}^{0}+b_{X}\left(\dmfifteen-1.1\right)\,.
\end{equation}
The details of this procedure are given in
Folatelli et al.\ (2008, in preparation).

In performing this calibration, it has become apparent that---for whatever
reason---a standard reddening law with the typical Milky Way value of
$\rvmw = 3.1$ does not provide a good fit to this sample of \sneia.
Instead, we obtain a value $\rvhost = 1.6\pm 0.1$
(Folatelli et al.\ 2008, in preparation).
Other groups have also found surprisingly low values of \rvhost\
\citep{Astier2006,Wang2006}.
This may be due to the \sneia\ being embedded in an environment that differs
significantly from a typical ISM, or perhaps there is another, intrinsic
color--magnitude effect at play that has been inadvertently combined
with $E(B-V)$.
Whatever the reason, we are interested in a distance determination and
therefore appeal to the empirical result $\rvhost = 1.6$.
However, we note that changing \rvhost\ in equation~(\ref{eq:lc_model}) does
not only affect the extinction term, but also changes the values of
$M_{X}^{0}$ because the calibrating sample has a non-zero average extinction.
Indeed, only a SN with an extinction equal to the average extinction of the
calibrating sample would be insensitive to changes in \rvhost.
In the case of SN 2007sr, the extinction is slightly larger than for the
calibrating sample, whence an increase in \rvhost\ would slightly decrease
the distance modulus.
Specifically, changing \rvhost\ from $1.6$ to $3.1$ would change the
estimated true distance modulus of this SN by $-$0.06 mag.

\begin{figure}
  \centering
  \includegraphics[scale=0.45]{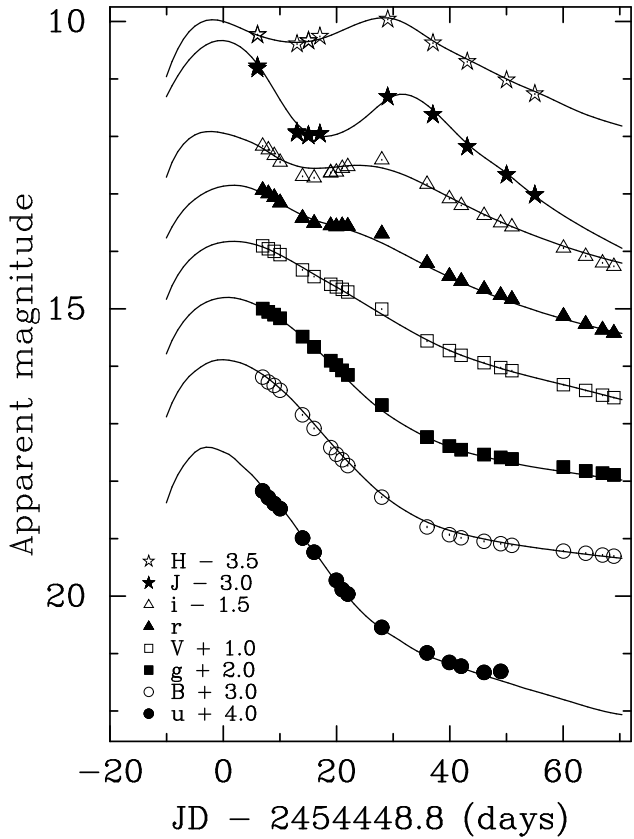}
  \caption{
Lightcurves in $u'Bg'Vr'i'JH$ for SN 2007sr.
For clarity, magnitudes in the different passbands have been offset by
amounts indicated in the legend.
{\em Lines} represent the best-fit templates.
Error bars are all smaller than the data points.
Epoch 0 corresponds to peak brightness in the $B$ band on 2007 December
14.3 UT.
  \label{fig02}}
\end{figure}

The light-curve templates $T_{X}$ are constructed in a manner similar
to that used by \citet{Prieto2006} and further described in
Burns et al.\ (2008, in preparation).
Essentially, the highest-quality optical light curves from the CSP,
which have clearly observed maxima, are fit with a cubic spline, from
which \dmfifteen, $t_{\rm max}$, and the peak magnitude are extracted.
These light curves then define a sparsely sampled 2D surface in 
\dmfifteen--$t$--$m_{X}$ space.
A template of any given \dmfifteen\ can then be generated by 2D interpolation
of this surface.
At present, there are an insufficient number of near-infrared (NIR) light
curves with which to construct templates in this way.
Fortunately, the decline-rate parameter for SN 2007sr ($\dmfifteen=0.97$ mag)
is nearly identical to that of SN 2006X ($\dmfifteen=0.98$ mag), for which
the CSP has obtained high-quality NIR photometry well before and after
$B$ maximum.
We therefore use spline fits of the $J$ and $H$ light curves of SN 2006X as
templates for fitting the $J$ and $H$ data of SN 2007sr.

The $u'Bg'Vr'i'JH$ light curves of SN 2007sr are shown in Fig.~\ref{fig02}
along with the best-fit models for the optical data given by
equation~(\ref{eq:lc_model}).
The extracted light-curve parameters of interest are the true distance
modulus of SN 2007sr, $\mu_0=31.74\pm0.07$ mag, the color excess due to
the host galaxy, $E(B-V)_{\rm host}=0.13\pm0.01$ mag, and the decline-rate
parameter $\dmfifteen=0.97\pm0.02$ mag.
The value of this parameter is a typical value for \sneia, and the
color excess is reasonably low as one would expect for a source far from
the host center.
The uncertainties are derived from the covariance matrix of the fit to
the light curves, scaled so that $\chi_{\nu}^{2}=1$.
To be conservative, we add to the uncertainty in $\mu_0$ the following
systematic errors:
(1) $\sigma_{M_X}=0.15$ mag for the intrinsic dispersion in \snia\
    luminosities,
(2) uncertainties in the calibration parameters of $\delta R_{V}=0.1$,
    $\delta M_{X}^{0}=0.04$ mag, and $\delta b_{X}=0.1$, and
(3) an uncertainty in the Hubble constant of $\delta H_0 = 8$ \kmsmpc.
With this error budget, we arrive at a final distance estimate to
SN 2007sr of $\dsnia=\DIa$ Mpc ($\dmodsnia=31.74\pm 0.27$ mag)
using the optical data.

As a check on this distance we also fitted the $J$- and $H$-band data with
templates generated from SN 2006X in order to estimate the peak magnitudes:
$m_{J}=13.31\pm0.06$ and $m_{H}=13.47\pm0.02$.  We then used the calibration
by \citet{Krisciunas2004}, $M_{J}=-18.57\pm0.14$ mag and $M_{H}=-18.24\pm0.18$
mag, to compute an average distance modulus of $\mu_0=31.80\pm0.11$ mag.
This modulus, based on NIR data and an independent calibration, is nearly
insensitive to assumptions about the reddening law of the dust in the host
galaxy and agrees with the formal modulus of $31.74\pm 0.07$ mag from the
optical data to within the combined errors, thus supporting the derived
distance of $\dsnia= \DIa$ Mpc.

\bigskip

\section{PROBLEMS WITH THE SHORT DISTANCE}
\label{sec3}

As we demonstrate below, the new \snia\ distance to \n4038/39 agrees well with
the distance inferred from the system's recession velocity and the
large-scale flow model by TBAD00.
If the short distance of $13.3\pm 1.0$ Mpc measured by \citet{savi08} were
to be correct, it would create at least three serious problems:
(1) \n4038/39 would have an exceptionally large peculiar recession velocity
of about $+$522 \kms;
(2) SN 2007sr would have had a peak luminosity differing by $\sim$\,1.1 mag,
or about 7\,$\sigma_{M_X}$, from the mean peak luminosity of \sneia; and
(3) The Antennae would lose their membership in Tully's (1988) Group 22--1
of 13 galaxies.
We now discuss these three problems in turn and then point out a likely
error in the derivation of the short distance by \citet{savi08}.

\subsection{Peculiar Recession Velocity}
\label{sec3.1}
\begin{figure}
  \centering
  \includegraphics[scale=0.58]{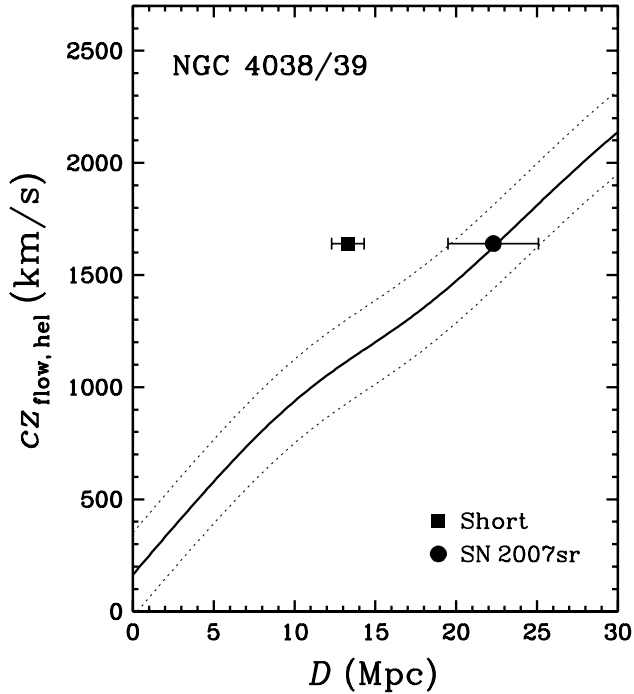}
  \caption{
Comparison of distances measured for \n4038/39 ({\em data points}) with
the large-scale flow model by TBAD00 ({\em solid line}).
Heliocentric recession velocities \czhel\ are plotted versus distance $D$.
Plotted at the new distance based on Type Ia SN 2007sr ({\em filled dot}),
the recession velocity of \n4038/39 falls well within 1\,$\sigma_{\rm th}$
({\em dotted lines}) of the large-scale flow, where $\sigma_{\rm th}$ is
the cosmic random radial velocity.
In contrast, when plotted at the short distance based on the TRGB
\citep[{\em square},][]{savi08}, the recession velocity of \n4038/39
lies 522 \kms\ or 2.8\,$\sigma$ above the flow.
For details, see \S~\ref{sec3.1}.
  \label{fig03}}
\end{figure}

A careful assessment of the many
radial velocities measured for \n4038/39 and compiled in the NASA/IPAC
Extragalactic Database (NED) yields a systemic heliocentric velocity of
$\czsyshel = 1640\pm 10$ \kms.
Figure~\ref{fig03} compares this systemic velocity with recession velocities
predicted by the large-scale flow model, which is based on surface-brightness
fluctuation distances to 300 early-type galaxies (TBAD00).
The Antennae are plotted at both the short distance and the new \snia\
distance, while the solid curve represents the predicted large-scale
flow velocity in their direction.
The undulation in the model curve reflects the gravitational influences
of the Virgo Cluster, whose center lies $32.2\degr$ away from The Antennae,
and of the Great Attractor.
As the figure illustrates, with the new \snia\ distance The Antennae fall
within $+15\,(\pm 188)$ \kms\ or $+$0.1\,\sigvelth\ of the predicted
flow velocity, where $\sigvelth = 187$ \kms\ is the best-fit ``thermal''
(or random) radial-velocity dispersion of the model.
If instead we choose the short distance, the peculiar velocity becomes
$+$522 ($\pm\,50$) \kms\ or $+$2.8\,\sigvelth, clearly an exceptionally
high value.
We conclude that the newly measured \snia\ distance is in significantly
better agreement with the best available large-scale flow model than the
short distance is.

\subsection{Peak Luminosity of SN 2007sr}
\label{sec3.2}
Our determination of a distance of $\DIa$ Mpc to SN 2007sr has been based
on the assumption that this SN reached the {\em mean} absolute magnitude
$M_X^0$ typical for its decline rate.
The quoted distance error includes a term reflecting the intrinsic
dispersion of the corrected peak absolute magnitudes $M_X$ of \sneia\
around the mean value $M_X^0$, which is $\sigma_{M_X} \approx 0.15$ mag
for CSP observations of similar quality (\S~\ref{sec2}).

\begin{figure}
  \centering
  \includegraphics[scale=0.34]{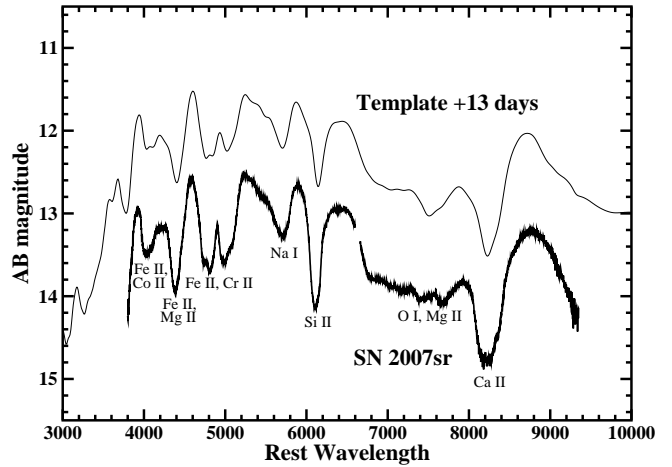}
  \caption{
Spectrum of SN 2007sr obtained with the Baade 6.5~m telescope and IMACS
spectrograph on 2007 December 27.33 UT, 13 days after $B$ maximum. 
Identifications of the major absorption features are taken from
\citet{bran08}.
The spectrum has been corrected for reddening produced in both the Milky
Way and The Antennae, and has been shifted to a rest wavelength scale.
The flux scale is plotted in AB magnitudes as defined by \citet{oke83}.
For comparison, the \snia\ template spectrum by \citet{hsia07} for an
epoch of +13~days is also shown.
  \label{fig04}}
\end{figure}

{\em If\,} the short distance of 13.3 Mpc were to be correct, SN 2007sr would
have reached a peak absolute magnitude that differed by
$\Delta(m-M)_0 = 5\log(22.3/13.3) = 1.12$ mag, or about 7\,$\sigma_{M_X}$,
from the mean value for \sneia.
This seems quite unlikely, given the spectral normality of SN 2007sr
\citep{nait07,umbr07}.
To illustrate this normality, Fig.~\ref{fig04} shows an optical spectrum
 obtained by the CSP on 2007 December 27.33 UT with the Baade 6.5~m
telescope and IMACS spectrograph, corresponding to an epoch of 13~days
after $B$ maximum.
For comparison, we include the SN~Ia template spectrum by \citet{hsia07}
at +13~days.
Although there are small differences between the two spectra, the overall
resemblance is extremely close, demonstrating that SN 2007sr was both
photometrically and spectroscopically a completely normal SN~Ia.
Hence, its extraordinarily low peak luminosity (by 7\,$\sigma_{M_X}$!),
were it to lie at 13.3 Mpc from us, is a second good reason for distrusting
the short distance.

\subsection{Group Membership}
\label{sec3.3}
Based on their position in the sky and
recession velocity, \citet{tull88} assigned The Antennae to Group 22--1 in
the Crater Cloud.
This group contains 13 galaxies, of which four---besides The Antennae---have
 measured distances.
None of these distances is less than 20 Mpc \citep{mado07}.

Two early-type galaxies of the group have had their distances measured via
both the surface-brightness fluctuation method and the Fundamental-Plane
method \citep{blak01,tonr01}.
The average distance for \n4024 (S0) is a rather uncertain $28.2\pm 4.3$ Mpc,
while that for \n4033 (E6) is a more concordant $21.7\pm 1.8$ Mpc.
Both galaxies lie within $\sim$\,1.0\degr\ from The Antennae.
Were The Antennae to lie at the short distance of 13.3 Mpc, it would be
difficult to understand why two nearby postulated group members with very
similar recession velocities ($+$52 and $-$24 \kms\ relative to The
Antennae, respectively) should lie at much larger distances.
Again, the evidence favors both the large-scale flow distance of
$\dflow = 22.5\pm 2.8$ Mpc (see \S~\ref{sec4}) and our new \snia\ distance
of $\dsnia = \DIa$ Mpc to The Antennae.

\begin{figure*}
  \centering
  \includegraphics[angle=-90,scale=0.70]{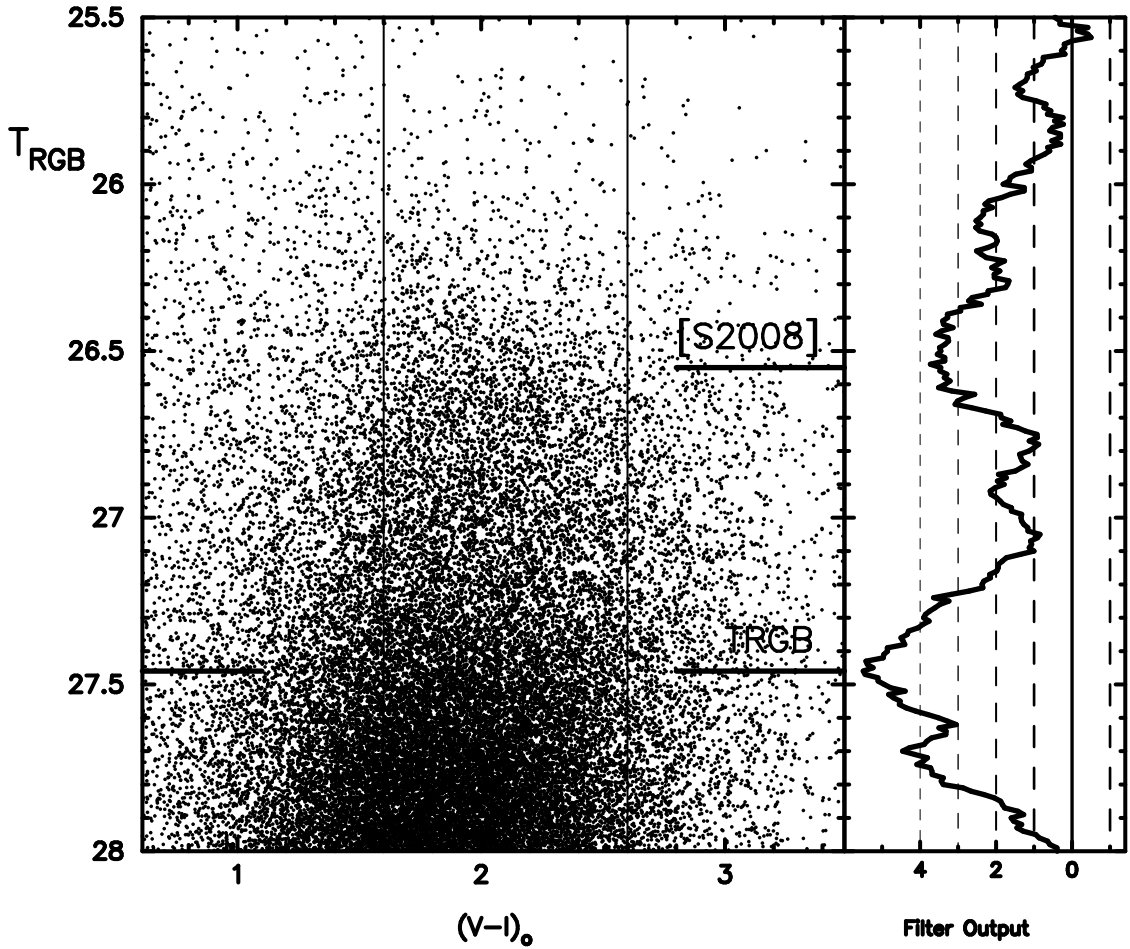}
  \caption{
({\em Left panel}) \trgbmag\ vs $(V-I)_0$ color--magnitude diagram for the
southern tidal-tail tip region of The Antennae; the \trgbmag\ magnitude
is corrected for Milky-Way foreground extinction and metallicity variations.
({\em Right panel}) Normalized Sobel filter response for all stars with
$1.6 \leq (V-I)_0 \leq 2.6$, plotted vs \trgbmag.
The response of highest relative significance is marked at
$\trgbmag = 27.46$ mag, with a reported significance of 5.4 sigma.
We identify this edge as the likely true TRGB, which lies about 0.9 mag below
the TRGB claimed by \citet{savi08}.
For details, see \S~\ref{sec3.4}.
  \label{fig05}}
\end{figure*}

\subsection{A Likely Misidentification of the TRGB}
\label{sec3.4}
Why is the TRGB distance of $13.3\pm 1.0$ Mpc determined by \citet{savi08}
so discrepant?
The answer seems to lie in a misidentification of what constitutes
the true TRGB of the mixed-age stellar populations near the tip of the
southern tidal tail.
The area imaged by \citeauthor{savi08}\ with \hst/ACS contains several
\hii\ regions \citep{schw78,mira92} and a very blue, dwarf-like feature
dubbed the ``S78 Region'' and now recognized to be the bent \hi-rich tip
of the southern tail \citep{hibb01}.

In their Figure~3, \citet{savi08} display two luminosity functions (LF)
for the RGB, one derived for the S78 Region and the other for a ``NE Region''
that is free of \hi\ and also, supposedly, of relatively young stars.
The LF of the S78 Region shows a relatively sharp rise at
$F814W_0 = 26.57$ mag (where $F814W$ is an instrumental magnitude close
to the Kron--Cousin $I$ magnitude), which \citeauthor{savi08}\ identify as
the TRGB but which---given the \hi\ and blue stellar content of the S78
Region---seems equally likely to be caused by relatively young AGB stars.
In contrast, the LF of the gas-free NE Region shows only a minor sharp
rise there, but features a more significant ``discontinuity at $F814W_0\approx
27.5$ mag'' \citep{savi08} that falls within $\la$\,0.2 mag of the TRGB
expected for a distance of 22 Mpc.
While \citeauthor{savi08}\ dismiss this discontinuity as possibly ``due to
the RGB of a younger, $\sim$\,200 Myr population,'' we suggest that
its stars---being {\em fainter} than the claimed TRGB---actually mark the
true location of the TRGB.

To check on this suggestion, we downloaded the ACS/WFC (Wide Field Camera)
frames obtained under proposal GO-10580 (P.I.: Ivo Saviane) from the \hst\
Archive, reprocessed them, and extracted a calibrated color--magnitude
diagram (CMD) for the entire tidal-tail tip region.
Our search and photometry algorithms, which use DOLPHOT
\citep[for technical details, see][]{mage08},
detected about 30,000 stellar objects brighter than $I\approx 28.0$ mag. 
Figure~\ref{fig05} (left panel) shows the CMD in a new form proposed by
\citet{mado08}, in which the traditional $I$ (or extinction-corrected $I_0$)
magnitude is replaced by the magnitude \trgbmag\ for the Red Giant Branch (RGB).
This magnitude is a color-corrected $I$-band magnitude specifically designed
to be insensitive to varying metallicity and is defined by
\begin{equation}
\trgbmag \equiv I_0 - \beta\,[(V-I)_0 - \gamma]\,,
   \label{eq:trgbdef}
\end{equation}
where the color slope $\beta= 0.20\pm 0.05$ is chosen to track the known,
metallicity-induced run of the $I_0$ magnitude of the TRGB as a function
of $(V-I)_0$, and $\gamma = 1.50$ mag is the fiducial color of reference
\citep{mado08}.
By construction, in any \trgbmag--$(V-I)_0$ diagram of an old stellar
population with a range of metallicities, the TRGB lies at some constant
value of \trgbmag.

To locate the TRGB objectively, we selected all stars in the color range
$1.6 \leq (V-I)_0 \leq 2.6$ and used a Sobel edge-detection kernel
\citep{lee93} to filter their numbers (in 0.03 mag bins) as a function of
\trgbmag.
The resulting filter output, expressed as a ratio between the filter
response and local Poisson noise, is shown in the right panel of
Fig.~\ref{fig05}.
In this panel, a peak at a normalized filter response of 2.0 indicates
an edge detection at a 2\,$\sigma$-level of significance.
Figure~\ref{fig05} shows a very clear, 5.4\,$\sigma$ detection of the TRGB
at\,\ $\trgbmag = 27.46\pm 0.12$ mag (marked ``TRGB''), or about 0.9~mag
{\em fainter} than the TRGB position of $I_0^{\rm TRGB}= 26.65$ claimed by
\citet{savi08}.
The latter position (at $\trgbmag\approx 26.55$) corresponds approximately
to the broad
$\sim$\,3.5\,$\sigma$ ``hump'' seen centered on $\trgbmag \approx 26.5$ mag
in the normalized filter response of Fig.~\ref{fig05}.
We conclude that \citet{savi08} probably misidentified youngish AGB stars
for the true TRGB, in the process overestimating its brightness by about
$-$0.9 mag.
This, then, is the likely source of their discrepant $13.3\pm 1.0$ Mpc
distance.

Our newly determined $\trgbmag = 27.46\pm 0.12$ mag yields a true distance
modulus to The Antennae of approximately\,\
$\dmodtrgb \equiv \trgbmag - M_I^{\rm TRGB} = 31.51\pm 0.17$ mag,
where $M_I^{\rm TRGB} = -4.05\pm 0.12$ mag is the absolute $I$ magnitude
for the TRGB of $\omega$~Cen \citep{bell04} and holds at the fiducial
color of reference \citep{mado08}.
This distance modulus, which is corrected for Milky Way foreground
extinction via \trgbmag\ [equation~(\ref{eq:trgbdef})], but not for
(varying) extinction within the tidal-tail tip region, corresponds to a
linear distance of $\dtrgb = 20.0\pm 1.6$ Mpc.
We postpone a more elaborate determination of \dmodtrgb, including a
careful assessment of local extinction and any possible small biases due
to the presence of young stellar populations, to a future paper dedicated
to that subject.
However, note that $\dtrgb = 20.0\pm 1.6$ Mpc agrees with both
$\dsnia = \DIa$ Mpc (\S~\ref{sec2}) and
$D_{\rm flow} = 22.5\pm 2.8$ Mpc (\S~\ref{sec4})
to within the combined errors, while it strongly disagrees with the
short distance of $13.3\pm 1.0$ Mpc derived by \citet{savi08}.
Thus, \dtrgb\ also supports the new \snia\ distance.

\section{CONCLUSIONS}
\label{sec4}

We have presented $u'Bg'Vr'i'JH$ light curves of SN 2007sr beginning 6 days
after $B$ maximum and obtained during four months by the Carnegie
Supernova Project.
Analysis of these light curves in conjunction with \sneia\ data from the
first-year campaign by the Project yields a formal true distance modulus
of $\mu_0=31.74\pm 0.07$ mag.
With all known systematic errors and uncertainties included, this distance
modulus becomes $\dmodsnia=31.74\pm 0.27$ mag, corresponding to a distance of
$\dsnia=\DIa$ Mpc to \n4038/39.
(Differential depth effects between SN 2007sr and the center of mass
of the galaxies are negligible).

This new \snia\ distance agrees well with the distance
$D_{\rm flow}= 22.5\pm 2.8$ Mpc estimated from the systemic recession
velocity of The Antennae and the large-scale flow model by TBAD00,
where the quoted error reflects the cosmic random radial velocity
($\sigvelth = 187$ \kms) of the model.
On the other hand, \dsnia\ disagrees strongly with the short distance
of $D = 13.3\pm 1.0$ Mpc estimated by \citet{savi08} from the
TRGB.
We have discussed three serious problems with such a short distance,
and have pointed out the likely misidentification of the TRGB by these
authors as the cause of the short distance.

Reprocessing their \hst/ACS frames and using an improved method of TRGB
detection, we have located what we believe to be the true TRGB at
$I_0 \approx \trgbmag = 27.46\pm 0.12$ mag, fully 0.9~mag below (i.e.,
fainter than) the TRGB claimed by \citeauthor{savi08}\ \
Analyzing photometry of this new, fainter TRGB we have derived a
preliminary distance of $\dtrgb = 20.0\pm 1.6$ Mpc.

Clearly, additional distance estimates to The Antennae (e.g., from Cepheids,
planetary nebulae, etc.) will be very valuable.
However, for the moment we see no reason for adopting any distance shorter
than 20 Mpc.
Given the concordant new $\dsnia = \DIa$ Mpc and recession-velocity
based $\dflow = 22.5\pm 2.8$ Mpc, both supported by our new, preliminary
$\dtrgb = 20.0\pm 1.6$ Mpc, we recommend using a conservative, rounded
value of $D = 22\pm 3$ Mpc as the best currently available distance to
The Antennae.

\acknowledgments

We thank David Murphy for his efforts in building instruments for the CSP,
John Tonry for his flow-model software and advice,
and Claudia Maraston for helpful discussions.
This research has made use of the NASA/IPAC Extragalactic Database,
which is operated by the Jet Propulsion Laboratory, California Institute
of Technology, under contract with NASA.
M.H.\ and G.F.\ ackowledge support from the Millennium Center for
Supernova Science through grant P06-045-F funded by ``Programa
Bicentenario de Ciencia y Tecnolog\'ia de CONICYT'' and ``Programa
Iniciativa Cient\'ifica Milenio de MIDEPLAN.''
M.H.\ acknowledges additional support from Centro de Astrof\'\i sica
FONDAP 15010003 and Fondecyt through grant 1060808.
Observations made under the Carnegie Supernova Project were supported in
part by the NSF through grants AST\,03-06969 and AST\,06-07438.

\bigskip

\appendix

\section{Photometry of SN 2007\lowercase{sr}}
\label{appA}

The magnitudes for SN 2007sr presented in this paper are on the CSP
natural system.
Details of the data reduction, calibration, and filter functions and color
terms defining this natural system can be found in \citet{Hamuy2006} and
% \citet{Contreras2008},
Contreras et al.\ (2008, in preparation),
as well as on the CSP website: \url{http://csp1.lco.cl/~cspuser1}.
In this Appendix, we briefly outline the natural system and present a
table of the photometry for SN 2007sr.

The $u'g'r'i'$ magnitudes are calibrated via the \citet{Smith2002} standard
stars.
A color sequence of these stars is used to establish color terms that
transform our instrumental magnitudes to the standard system of
\citet{Smith2002}.
However, since \sneia\ have spectra that are very different from those of
normal stars and evolve with age, these color terms cannot be used to
transform the SN magnitudes to the standard system.
Instead, we use the color terms in reverse to transform the standard
magnitudes to our natural system, and then calibrate our SN observations
using these natural magnitudes.
It is in this system that we present the photometry of SN 2007sr and
perform our entire analysis.

The $BV$ magnitudes are treated in a similar way, though using the system
of \citet{Landolt1992} standard stars.
The $JH$ magnitudes are calibrated using the \citet{Persson1998} standards,
for which the instrument and filters used were nearly identical to those
employed by the CSP and, therefore, were equivalent to our natural system.

Table~\ref{tab:photometry} presents the photometry for SN 2007sr in the
CSP natural system.
The listed uncertainties include the usual photon statistics as 
well as errors in the zero points.

%%%%%%%%%%%%%%%%%%%%%%%  Table 1  %%%%%%%%%%%%%%%%%%%%%%%%
%%%%%%%  (This placement only for `emulateapj')  %%%%%%%%%
\clearpage
\begin{deluxetable}{lcccccccc}
\tablecolumns{9}
\tablewidth{0pc}
\tablecaption{CSP Photometry for SN 2007\lowercase{sr}\label{tab:photometry}}
\tablehead{\colhead{JD $-$ 2,453,000} & \colhead{$u^\prime$} & 
           \colhead{$g^\prime$} & \colhead{$r^\prime$} &
          \colhead{$i^\prime$} & \colhead{$B$} & \colhead{$V$} &
          \colhead{$J$} & \colhead{$H$}}
%% \rotate
\startdata
1454.84 & \ldots& \ldots& \ldots& \ldots& \ldots& \ldots& 13.776(0.019)& 13.723(0.005)\\
1455.76 & 14.133(0.013)& 12.999(0.004)& 12.924(0.005)& 13.664(0.006)& 13.165(0.005)& 12.856(0.005)& \ldots& \ldots\\
1456.76 & 14.241(0.014)& 13.051(0.004)& 12.983(0.006)& 13.733(0.007)& 13.251(0.007)& 12.904(0.006)& \ldots& \ldots\\
1457.80 & 14.345(0.009)& 13.101(0.004)& 13.047(0.005)& 13.826(0.005)& 13.310(0.006)& 12.947(0.005)& \ldots& \ldots\\
1458.84 & 14.429(0.012)& 13.168(0.008)& 13.136(0.009)& 13.942(0.010)& 13.390(0.007)& 13.000(0.008)& \ldots& \ldots\\
1461.82 & \ldots& \ldots& \ldots& \ldots& \ldots& \ldots& 14.912(0.017)& 13.878(0.004)\\
1462.80 & 14.936(0.010)& 13.484(0.005)& 13.406(0.005)& 14.190(0.006)& 13.810(0.006)& 13.265(0.006)& \ldots& \ldots\\
1463.84 & \ldots& \ldots& \ldots& \ldots& \ldots& \ldots& 14.971(0.006)& 13.837(0.004)\\
1464.84 & 15.180(0.012)& 13.665(0.007)& 13.498(0.007)& 14.216(0.009)& 14.040(0.008)& 13.389(0.005)& \ldots& \ldots\\
1465.87 & \ldots& \ldots& \ldots& \ldots& \ldots& \ldots& 14.959(0.007)& 13.767(0.007)\\
1467.77 & \ldots& 13.910(0.006)& 13.540(0.004)& 14.130(0.004)& 14.367(0.007)& 13.537(0.006)& \ldots& \ldots\\
1468.77 & 15.662(0.018)& 13.986(0.006)& 13.556(0.006)& 14.107(0.007)& 14.470(0.006)& 13.591(0.006)& \ldots& \ldots\\
1469.80 & 15.819(0.016)& 14.083(0.005)& 13.544(0.004)& 14.049(0.006)& 14.571(0.006)& 13.640(0.006)& \ldots& \ldots\\
1470.79 & 15.898(0.017)& 14.161(0.006)& 13.552(0.006)& 14.021(0.006)& 14.671(0.007)& 13.686(0.006)& \ldots& \ldots\\
1471.87 & \ldots& \ldots& \ldots& \ldots& \ldots& \ldots& 14.708(0.007)& 13.534(0.007)\\
1476.78 & 16.478(0.013)& 14.696(0.006)& 13.690(0.004)& 13.906(0.006)& 15.206(0.006)& 14.011(0.006)& \ldots& \ldots\\
1477.84 & \ldots& \ldots& \ldots& \ldots& \ldots& \ldots& 14.313(0.006)& 13.469(0.005)\\
1484.78 & 16.924(0.029)& 15.252(0.008)& 14.206(0.007)& 14.331(0.009)& 15.728(0.010)& 14.574(0.010)& \ldots& \ldots\\
1485.82 & \ldots& \ldots& \ldots& \ldots& \ldots& \ldots& 14.630(0.009)& 13.880(0.008)\\
1488.73 & 17.086(0.048)& 15.407(0.008)& 14.429(0.006)& 14.583(0.007)& 15.863(0.011)& 14.734(0.006)& \ldots& \ldots\\
1490.79 & 17.150(0.013)& 15.466(0.005)& 14.517(0.004)& 14.699(0.004)& 15.917(0.008)& 14.814(0.005)& \ldots& \ldots\\
1491.88 & \ldots& \ldots& \ldots& \ldots& \ldots& \ldots& 15.186(0.009)& 14.205(0.007)\\
1494.85 & 17.260(0.015)& 15.553(0.007)& 14.657(0.006)& 14.873(0.008)& 15.983(0.009)& 14.941(0.007)& \ldots& \ldots\\
1497.78 & 17.251(0.035)& 15.603(0.006)& 14.762(0.006)& 15.004(0.007)& 16.027(0.009)& 15.025(0.006)& \ldots& \ldots\\
1498.82 & \ldots& \ldots& \ldots& \ldots& \ldots& \ldots& 15.674(0.011)& 14.527(0.008)\\
1499.77 & \ldots& 15.630(0.005)& 14.830(0.005)& 15.074(0.006)& 16.057(0.007)& 15.076(0.005)& \ldots& \ldots\\
1503.82 & \ldots& \ldots& \ldots& \ldots& \ldots& \ldots& 16.021(0.013)& 14.764(0.008)\\
1508.86 & \ldots& 15.768(0.006)& 15.123(0.007)& 15.444(0.008)& 16.164(0.010)& 15.312(0.006)& \ldots& \ldots\\
1511.79 & \ldots& \ldots& \ldots& \ldots& \ldots& \ldots& 16.527(0.016)& 15.083(0.012)\\
1512.79 & \ldots& 15.834(0.006)& 15.261(0.007)& 15.585(0.008)& 16.207(0.009)& 15.407(0.007)& \ldots& \ldots\\
1515.76 & \ldots& 15.870(0.007)& 15.360(0.005)& 15.700(0.007)& 16.237(0.009)& 15.490(0.007)& \ldots& \ldots\\
1517.76 & \ldots& 15.896(0.006)& 15.417(0.006)& 15.759(0.007)& 16.260(0.010)& 15.530(0.007)& \ldots& \ldots\\
1523.79 & \ldots& 15.996(0.006)& 15.620(0.005)& 15.973(0.007)& 16.349(0.008)& 15.686(0.006)& \ldots& \ldots\\
1527.77 & \ldots& 16.047(0.006)& 15.737(0.008)& 16.115(0.010)& 16.396(0.009)& 15.784(0.008)& \ldots& \ldots\\
1532.70 & \ldots& \ldots& \ldots& \ldots& \ldots& 15.915(0.020)& \ldots& \ldots\\
1538.75 & \ldots& 16.214(0.005)& 16.089(0.004)& 16.470(0.005)& 16.571(0.006)& 16.051(0.004)& \ldots& \ldots\\
1540.68 & \ldots& 16.218(0.006)& 16.131(0.008)& 16.521(0.009)& 16.607(0.010)& 16.140(0.012)& \ldots& \ldots\\
1542.66 & \ldots& 16.276(0.007)& 16.217(0.008)& 16.612(0.010)& 16.633(0.010)& 16.138(0.008)& \ldots& \ldots\\
1549.71 & \ldots& 16.396(0.009)& 16.447(0.011)& 16.819(0.014)& 16.725(0.019)& 16.298(0.011)& \ldots& \ldots\\
1558.65 & \ldots& 16.531(0.006)& 16.723(0.008)& 17.098(0.011)& 16.885(0.012)& 16.483(0.007)& \ldots& \ldots\\
1566.76 & \ldots& 16.663(0.007)& 16.964(0.008)& 17.305(0.011)& 17.017(0.012)& 16.648(0.009)& \ldots& \ldots\\
1571.66 & \ldots& 16.747(0.007)& 17.096(0.009)& 17.422(0.013)& 17.099(0.012)& 16.750(0.009)& \ldots& \ldots\\
1578.70 & \ldots& 16.870(0.010)& 17.291(0.008)& 17.600(0.009)& 17.208(0.009)& 16.935(0.059)& \ldots& \ldots\\
\enddata
\end{deluxetable}

\clearpage

\end{document}